\documentclass[12pt]{article}

\usepackage{newtxtext,newtxmath}
\usepackage{graphicx}

\usepackage[letterpaper,margin=1in]{geometry}
\usepackage{tabularx} % add in preamble
\usepackage{makecell}

\renewenvironment{abstract}
	{\quotation}
	{\endquotation}

\date{}

\makeatletter
\renewcommand{\fnum@figure}{\textbf{Figure \thefigure}}
\renewcommand{\fnum@table}{\textbf{Table \thetable}}
\makeatother

\usepackage{scicite}

\usepackage{url}

\def\scititle{
	Advanced Control of Electron Beams: Tailoring X-ray Production with Programmable Laser Shaping
}
\title{\bfseries \boldmath \scititle}

\author{
Jack~Hirschman$^{1,2\ast}$,\and
Randy~Lemons$^{2}$,\and
Hao~Zhang$^{2,3}$,\and
Razib~Obaid$^{2}$,\and
River~Robles$^{1,2}$,\and
Paris~Franz$^{2}$,\and
Benjamin~Mencer$^{2}$,\and
Nicole~Neveu$^{2}$,\and
Matthew~Britton$^{2}$,\and
David~Cesar$^{2}$,\and
Nicolas~Sudar$^{2}$,\and
Zhen~Zhang$^{2}$,\and
Justin~Baker$^{4}$,\and
Chad~Pennington$^{3}$,\and
Kurtis~Borne$^{2,5}$,\and
Taran~Driver$^{2}$,\and
Kirk~A.~Larsen$^{2}$,\and
Veronica~Guo$^{1,2}$,\and
Yuantao~Ding$^{2}$,\and
Gabriel~Just$^{2}$,\and
Feng~Zhou$^{2}$,\and
James~Cryan$^{2}$,\and
Joseph~Robinson$^{2}$,\and
Ryan~Coffee$^{2}$,\and
Agostino~Marinelli$^{2}$,\and
and Sergio~Carbajo$^{1,2,3,6,7}$
\and
\small$^{1}$Department of Applied Physics, Stanford University, Stanford, CA 94305, USA.
\and
\small$^{2}$SLAC National Accelerator Laboratory, Menlo Park, CA 94025, USA.
\and
\small$^{3}$Department of Electrical \& Computer Engineering, UCLA, Los Angeles, CA 90095, USA.
\and
\small$^{4}$Department of Mathematics, UCLA, Los Angeles, CA 90095, USA.
\and
\small$^{5}$J.R. Macdonald Laboratory, Department of Physics, Kansas State University, Manhattan, KS 66506, USA.
\and
\small$^{6}$Physics \& Astronomy Department, UCLA, Los Angeles, CA 90095, USA.
\and
\small$^{7}$California NanoSystems Institute, Los Angeles, CA 90095, USA.
\and
\small$^\ast$Corresponding author. Email: jhirschm@stanford.edu}

\begin{document} 

% Insert the title and author list
\maketitle

% Abstract, in bold
% There are strict length limits, and not all formats have abstracts.
% Consult the journal instructions to authors for details.
% Do not cite any references in the abstract.
\begin{abstract} \bfseries \boldmath
% Start with one or two sentences of background
Leveraging the full scientific capabilities of next-generation high-repetition-rate free-electron lasers requires programmable control over electron-beam properties at their source. 
The photoinjector drive laser defines the electron beam's initial six-dimensional phase-space distribution, yet has historically been limited to Gaussian or static flat-top profiles, with most manipulation occurring downstream. 
Here we demonstrate software-programmable ultraviolet pulse shaping at the LCLS-II photoinjector as a source-level actuator that complements traditional accelerator controls. 
Using a coupled architecture combining dispersion-controlled nonlinear frequency conversion with spatial-light-modulator spectral shaping, we generate user-defined temporal structures and observe their imprint on electron bunches through high-resolution time-domain diagnostics.
Laser-imposed multi-peaked modulation persists through acceleration, magnetic compression, and undulator transport with shot-to-shot repeatability, producing clearly resolved current structure in the compressed beam. 
Variance-based reconstruction from transverse deflecting cavity measurements reveals structured X-ray emission profiles exhibiting temporal features consistent with the programmed laser waveform.
By providing rapid, software-controlled reconfiguration of electron-beam initial conditions, this source-level control approach establishes a programmable upstream actuator for future adaptive optimization and autonomous facility operation at high-repetition-rate light sources.
\end{abstract}
\section{Introduction}
Electrons are the most versatile charged-particle sources for radiation generation, underpinning an extraordinary range of light sources from compact laboratory-scale systems to kilometer-scale synchrotrons and free-electron lasers (FELs)~\cite{esarey2009physics,schoenlein2019recent,pellegrini2016physics,zhang2025charged}.
By controlling their phase space and trajectories, the properties of the radiation they emit can be tailored across a broad parameter space~\cite{ha2022bunch,pellegrini2016physics,loisch2018photocathode}.
When decelerated in dense materials, electrons generate broadband bremsstrahlung; when guided through magnetic lattices, they emit intense synchrotron and undulator radiation; and when scattered from intense laser pulses, they produce tunable Compton or inverse Compton X-rays~\cite{bethe1934stopping,wiedemann2003synchrotron,schwinger1949classical,petrillo2023state,wong2024free,wong2021control}.
In all cases, the emitted radiation is governed by the electron phase space and its coupling to electromagnetic fields, making phase-space control a unifying principle for programmable radiation generation.

These electron-driven sources have had wide--ranging scientific, medical, and industrial impact, from conventional linac-based X-ray systems to emerging compact accelerator and laser–electron scattering platforms~\cite{schoenlein2019recent,amaldi1999cancer,kutsaev2025small,barty2024,reutershan2025scanning}.
Recent advances in laser- and plasma-based acceleration have further demonstrated that GeV-class electron beams and associated secondary radiation can be generated in compact geometries, highlighting the growing convergence of accelerator physics and ultrafast photonics~\cite{esarey2009physics,albert2023principles}.
Together, these developments underscore the increasing importance of precise, flexible control over electron–photon interactions as a pathway toward next-generation radiation sources.

At the facility scale, this convergence reaches its most advanced realization in x-ray free-electron lasers (XFELs), which combine ultrahigh brightness, femtosecond and sub-femtosecond pulse durations, and atomic-scale spatial coherence~\cite{huang2007review}.
Facilities such as the Linac Coherent Light Source (LCLS) and its superconducting upgrade, LCLS-II, have established new performance frontiers through ultralow-emittance photoinjectors, high-gradient acceleration, and increasingly sophisticated timing and diagnostic infrastructure, providing a platform in which source-level control of the electron beam directly governs the properties of the emitted x-rays~\cite{PhysRevAccelBeams.24.073401, zhang2024linac, frisch2014electron, walter2021multi,PhysRevAccelBeams.24.073401,brachmann2024commissioning}.

Crucially, the transition from 120-Hz to MHz-class operation represents an independent and transformative shift, producing data volumes orders of magnitude larger than in previous generations and fundamentally changing how experiments must be conducted and optimized~\cite{thayer2024massive,rahimifar2024accelerating}.
If diagnostics can operate commensurately, this regime enables not only high-throughput measurements, but also new operational paradigms--including multiplexed beam delivery to multiple end stations, rapid reconfiguration of beam properties between experiments, and adaptive tuning of the electron beam during runtime based on real-time feedback from single-shot X-ray diagnostics~\cite{walter2021multi, hirschman2022edge, hirschman2025hybrid, scheinker2018demonstration,li2024_carte, brachmann2024commissioning, heimann2025characterization, dolgashev24lcls, dolgashev2022new}.
In this context, offline analysis remains essential for comprehensive interpretation, but cannot practically provide the low-latency feedback required to steer experiments shot-by-shot or in real time at MHz repetition rates; unused or suboptimally configured shots therefore represent a direct loss of scientific opportunity.

In parallel, recent advances in machine-learning-assisted optimization and model-independent feedback control have demonstrated the potential for autonomous beamline tuning, leveraging real-time electron-beam and photon diagnostics to maximize XFEL output~\cite{scheinker2019model, scheinker2018demonstration, edelen2020machine, roussel2021multiobjective, hirschman2025hybrid, mishra2025start}.
Building on these efforts, the integration of digital-twin frameworks has emerged as a powerful path toward end-to-end accelerator optimization, unifying source-level models, beam dynamics, and photon diagnostics within a single predictive architecture~\cite{dorigo2023toward, biedron2022snowmass21, mishra2025start, edelen2024machine, hirschman:ipac25-mopb040}.
Together, these developments highlight a growing bottleneck: while downstream diagnostics, modeling, and optimization capabilities have advanced rapidly, the set of fast, flexible upstream control knobs remains comparatively constrained~\cite{kling2024roadmap, doe2023laserbrn}.

This confluence of high repetition rate, high data throughput, and emerging automation places a premium on source-level actuators that can operate on relevant timescales, beyond conventional accelerator tuning elements that are limited in bandwidth or operational agility.
In particular, programmable control of the photoinjector drive laser--which defines the electron beam’s initial six-dimensional phase-space distribution--offers a uniquely powerful lever for shaping the longitudinal phase space of the electron beam and, ultimately, the temporal structure of the emitted x-rays~\cite{ha2022bunch,carbajo2025structured,luiten2004realize}.
Bridging this gap from downstream optimization to source-level programmability is essential to realize the next generation of fully adaptive, digital-twin-enabled XFEL facilities.

Historically, temporal shaping at the injector has been implemented using passive techniques such as pulse stacking with birefringent crystals, enabling near-flattop current profiles for emittance compensation and tailored ramped structures for wakefield-driven acceleration~\cite{10.1063/1.3080991}.
Owing to the fast response of modern photocathode materials--such as cesium telluride (CsTe)--, the emitted electron current roughly follows the temporal intensity profile of the drive laser, enabling rapid imprinting of laser-defined temporal structure onto the emitted electron bunch, which subsequently evolves under injector space-charge dynamics~\cite{ha2022bunch, Monaco2022Photocathode, loisch2022}.
While reliable and widely adopted, these interferometric stacking methods inherently produce residual temporal microstructure and intensity ripples arising from discrete pulse replicas, which can introduce unwanted current modulation and limit precise phase-space control~\cite{lemons2022temporal}.
More advanced active shaping approaches have subsequently been developed, including programmable spatiotemporal control of the photocathode laser to directly engineer longitudinal and transverse beam structure for low-emittance, high-brightness FEL injectors~\cite{ilia:ipac2025-thpb027, Xu:IPAC2019-TUPTS104}. 
Beyond emittance optimization, structured optical modulation of the injector laser has recently been leveraged to impose periodic energy and density microstructures on relativistic electron beams, enabling programmable spectral tailoring of free-electron-laser radiation, including continuous tunability across the terahertz regime~\cite{Kang2026}.

The long-term objective of next-generation XFEL operation is full spatiotemporal control of the photoinjector drive laser~\cite{ha2022bunch, carbajo2025structured}.
Such control extends beyond conventional flattop shaping for emittance reduction, unlocking and complementing advanced operating modes that would otherwise require slower or downstream modulation techniques, including controlled microbunching, seeded operation, or dual-pulse generation for X-ray pump–X-ray probe experiments~\cite{Zhang2020-av, li2009laser, yang2002low, marinelli2016optical, robles2025spectrotemporal, li2024_carte, hemsing2014beam, guo2024experimental}.
When combined with high-throughput, single-shot diagnostics and real-time data processing, source-level programmability provides a foundation for future adaptive and data-driven XFEL operation, in which beam properties can be modified during runtime in response to experimental needs.
Looking ahead, increased control at the source may further enable deterministic access to emerging XFEL mode-locking concepts~\cite{hemsing2014beam, hu2025, thompson2008mode}, which in turn could open new opportunities for nuclear photonics through X-ray frequency-comb heterodyning and dual-comb spectroscopic measurements—capabilities that are presently restricted to the optical domain~\cite{burvenich2006nuclear,Heeg2021CoherentNuclearExcitons, liao2011nuclear}.
More broadly, these developments complement a growing ecosystem of advanced X-ray pulse-shaping techniques aimed at extending both the temporal and spectral structure of XFEL output~\cite{robles2025spectrotemporal, PhysRevLett.115.114801, Maroju2020}.

Photocathode drive-laser shaping represents one powerful and direct strategy for achieving longitudinal control of the electron beam at the source.
While this approach is inherently limited to accelerator platforms based on photoemission, it offers a uniquely upstream and broadly compatible control mechanism that can address many of the performance demands anticipated for future XFEL facilities and science cases. 
Prior studies have demonstrated early progress toward tailored pulse envelopes and static shaping approaches; however, adaptable and fully programmable source-level control remains largely unexplored at operating FEL facilities~\cite{yang2002low, lemons2025nonlinear, NEVEU2025170065}.

Here we demonstrate, for the first time, programmable shaping of X-ray pulses via direct modulation of the ultraviolet photoinjector drive laser beyond the temporal flattop.
We introduce a coupled shaping architecture that integrates a novel upconversion scheme--Dispersion-Controlled Nonlinear Synthesis (DCNS)~\cite{lemons2022temporal, lemons2025nonlinear}--with an upstream, pre-amplifier spatial light modulator (SLM) for programmable spectral shaping.
Implemented on the LCLS-II photoinjector laser, this system establishes a direct, source-level link between laser waveform design and electron-beam longitudinal control, ultimately enabling tailored X-ray pulse structures at high repetition rate.

In this work, we operate the DCNS in a static conversion mode and vary the upstream SLM mask to generate several distinct ultraviolet temporal profiles. Three representative cases are presented. 
The UV pulses are characterized using an in-line cross-correlator, and the resulting electron beams are diagnosed both early in the accelerator and after the undulators using transverse-deflecting cavities (TCAVs). 
These measurements provide direct, proof-of-principle evidence that programmable modulation of the photoinjector laser imprints measurable structure on the electron beam and the emitted X-ray pulses, establishing a foundation for adaptive, source-level control in next-generation, high-rate FEL facilities.

\section{Results and Discussion}

\begin{figure}[ht]
\centering
\includegraphics[width=\linewidth]{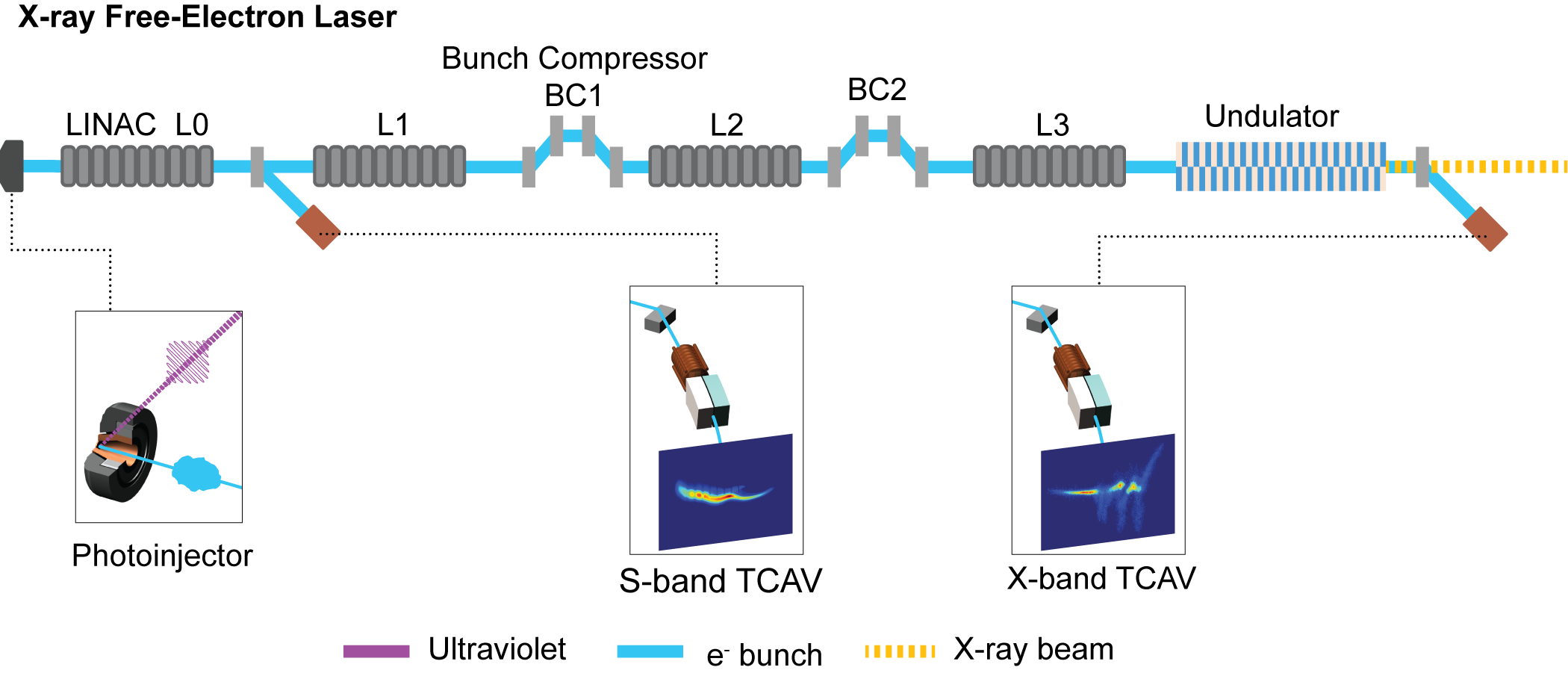}
\caption{
\textbf{LCLS-II X-ray production beamline overview.}
At the photoinjector, an ultraviolet laser pulse illuminates the photocathode to generate an electron bunch. 
The bunch is accelerated in the LINAC sections (L0–L3) and compressed in magnetic chicanes (BC1 and BC2) before producing X-rays in the undulator. 
Electron phase space is diagnosed immediately after the photoinjector using the S-band TCAV and downstream of the undulator using the X-band TCAV.
}
\label{fig:beamline}
\end{figure}

\subsection{Photoinjector Laser and Beamline}
The LCLS-II produces X-rays through a cascaded series of nonlinear operations (shown in Fig.~\ref{fig:beamline}), beginning with the photoinjector, where an ultraviolet (UV) pulse liberates an electron distribution from a CsTe photocathode in the normal-conducting, continuous wave (CW) mode, very-high frequency (VHF)-band radio frequency (RF) injector gun~\cite{Mondal2025, 10.1007/978-3-030-01629-6_24, zhang2024linac}.
The LCLS-II CsTe photocathode emission response time is well below the picosecond scale, ensuring that the applied ultraviolet laser pulse shaping is mapped into the emitted electron bunch profile~\cite{Monaco2022Photocathode, loisch2022}.
A buncher cavity and focusing solenoids downstream of the gun provide initial longitudinal compression and transverse matching of the beam. 
These electrons are then accelerated through a series of superconducting accelerator sections. Along the beamline, two magnetic bunch compressors (BC) sections (BC1 and BC2)--located downstream of the L1 and L2 accelerator sections--progressively shorten the electron bunch to achieve the peak currents required for lasing.
The fully accelerated and compressed beam is then transported to the soft X-ray (SXR) undulator line, where it radiates coherently to produce intense, femtosecond X-ray pulses.
Additional details of the injector and beamline layout are provided in the Methods Section~\ref{sec:methods}.

\begin{figure}[ht]
\centering
\includegraphics[width=\linewidth]{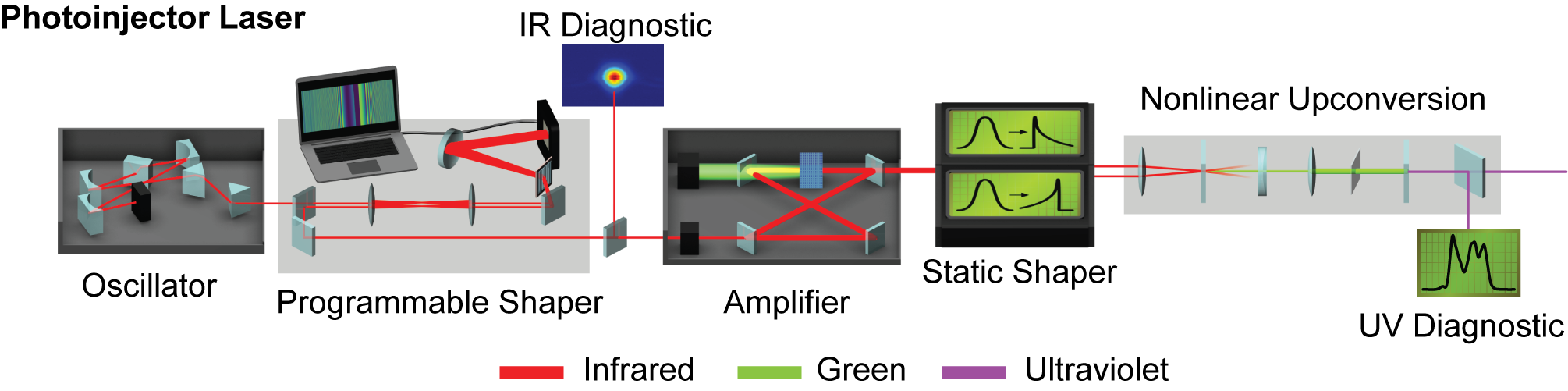}
\caption{
\textbf{R\&D Photoinjector Laser.}
The photoinjector laser comprises an oscillator, a programmable SLM-based spectral shaper, amplifier, compressor and stretcher static shapers, and noncollinear SFG and SHG for upconversion to UV, along with a FROG IR diagnostic and UV cross-correlator. 
}
\label{fig:laser}
\end{figure}

The initial photoinjector UV laser has multiple configuration modes depending on facility operation.
For standard operation, a high-power infrared (IR) drive laser followed by two stages of second-harmonic generation (SHG) produces a $\sim$20 ps Gaussian UV pulse. 
In the X-ray Laser Enhanced Attosecond Pulse (XLEAP) mode, an additional compressed IR laser ($\sim$1 ps Gaussian pulse) also undergoes two SHG stages and is temporally overlapped with the original drive UV pulse, introducing an additional modulation that seeds microbunching in the electron bunch for isolated attosecond X-ray pulse production~\cite{zhang2024linac, duris2020tunable, zholents2005method}.

Here, we experimentally demonstrate for the first time an adaptable laser setup at the LCLS-II photoinjector (Fig.~\ref{fig:laser}).
Specifically, the standard SHG upconversion stages are replaced with Dispersion-Controlled Nonlinear Synthesis--a technique that applies equal and opposite second-order (SOD) and third-order (TOD) dispersion to two IR pulses post-amplification and performs noncollinear sum-frequency generation (SFG) followed by SHG, producing a relatively flat-top UV temporal baseline profile between 15 ps and 20 ps~\cite{lemons2022temporal, lemons2025nonlinear, zhang2026upstreamlaserbasedlongitudinalenhancement}.
To make this setup programmable, a liquid crystal SLM in a 4f configuration is incorporated downstream of a Light Conversion Flint oscillator and upstream of a 40W Light Conversion Carbide regenerative amplifier to perform spectral shaping. 
Further details on the shaper implementation, dispersion tuning, and upconversion configuration are provided in the Methods section, with the complete photoinjector laser chain shown in Fig.~\ref{fig:laser}.

The UV temporal profile is diagnosed using a cross-correlator, and the generated electron bunch is diagnosed early in the accelerator by an S-band transverse deflecting cavity (STCAV) and after the undulators by an X-band transverse deflecting cavity (XTCAV) (Fig.~\ref{fig:beamline})~\cite{Behrens2014, ding2011, PhysRevSTAB.13.092801, PhysRevSTAB.3.032801, Emma2000TransverseRF}.
These electron diagnostics provide time-resolved measurements of the longitudinal electron bunch profile and an approximation of X-ray temporal pulse emission, enabling correspondence between applied laser shaping and downstream beam dynamics.
Additional details on diagnostics are provided in the Methods Section~\ref{sec:methods}.

\subsection{UV Temporal Profiles, Electron Bunch Imaging \& X-ray Generation}

\begin{figure}[htpb]
\centering
\includegraphics[width=\linewidth]{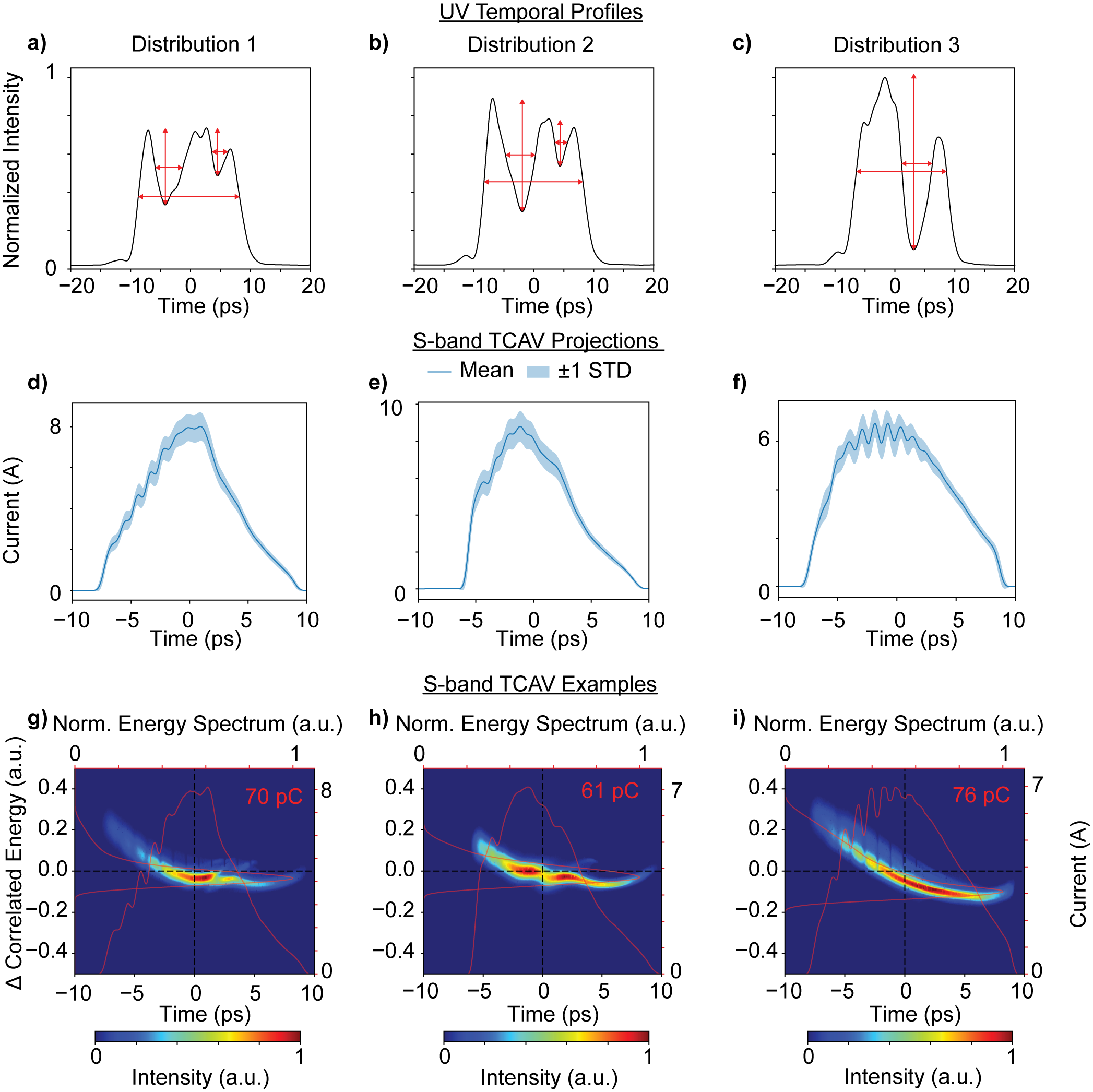}
\caption{
\textbf{Early electron beam generation.}
(a–c) UV photoinjector laser temporal profiles used to generate the three electron-beam distributions: shallow three-hump modulation (distribution 1), deep three-hump modulation (distribution 2), and deep two-hump modulation (distribution 3). 
(d–f) Mean longitudinal current profiles measured using the pre-undulator S-band TCAV for each distribution, with shaded regions indicating the shot-to-shot standard deviation. 
(g–i) Representative S-band TCAV phase-space spectrograms corresponding to the measured current projections. 
The vertical axis represents the correlated energy deviation of the electron beam, including contributions from energy chirp and compression.
}
\label{fig:stcav}
\end{figure}

\begin{figure}[htpb]
\centering
\includegraphics[width=\linewidth]{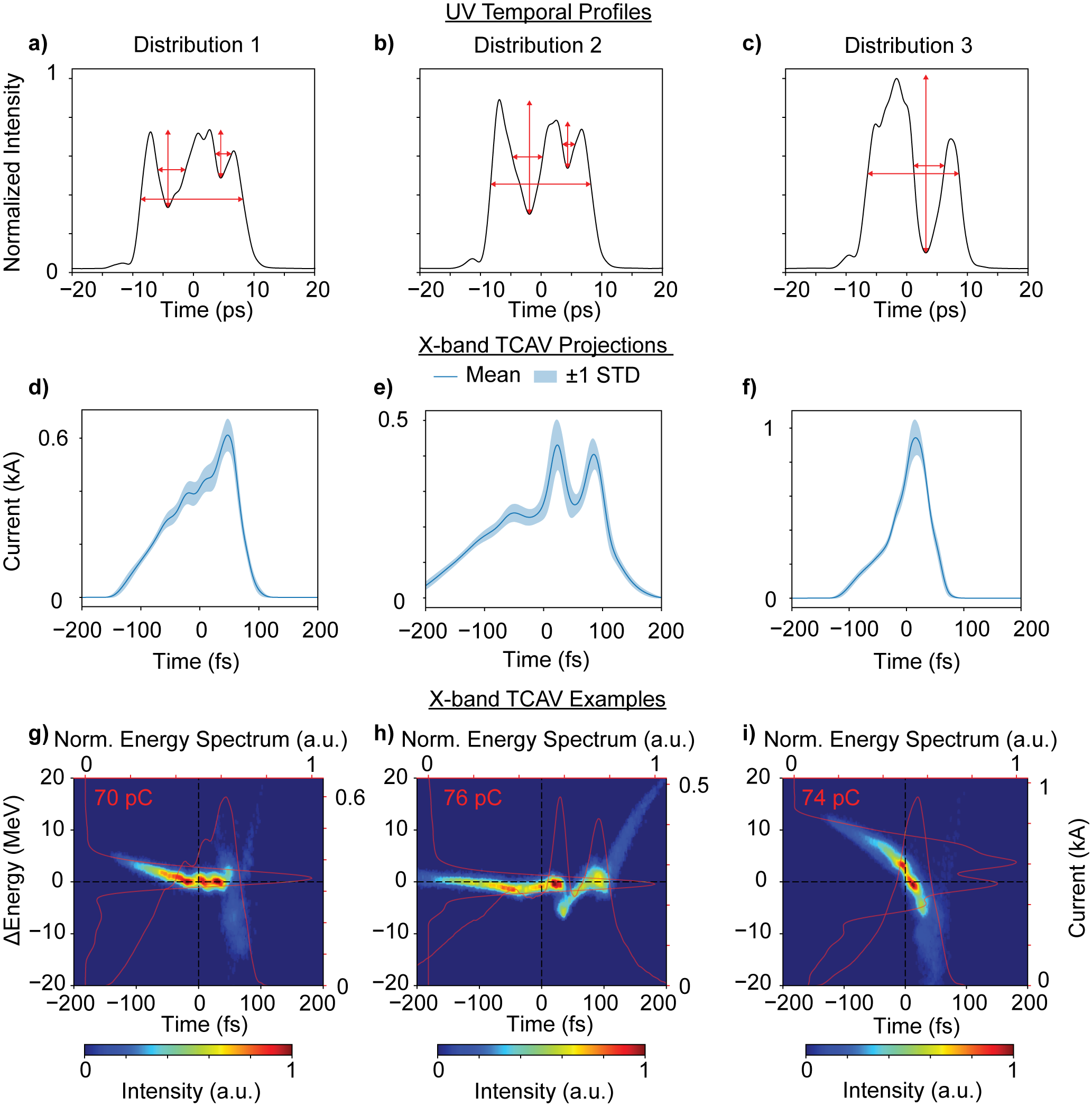}
\caption{
\textbf{UV and downstream electron beam measurements.}
(a–c) Ultraviolet photoinjector laser temporal profiles for the three shaped distributions, shown again for comparison with downstream measurements. 
(d–f) Mean longitudinal current profiles measured using the post-undulator X-band TCAV for each distribution, with shaded regions indicating the shot-to-shot standard deviation. 
(g–i) Representative X-band TCAV phase-space spectrograms from which the current projections are obtained.
}
\label{fig:xtcav}
\end{figure}

To demonstrate the robustness of the programmable photoinjector-laser shaping platform for tailored X-ray generation, we explore a series of UV temporal profiles with progressively deeper modulations imposed on a flat-top baseline.
Three representative cases were implemented during beam time: a shallow three-hump (distribution 1), a deep three-hump (distribution 2), and a deep two-hump modulation (distribution 3) shown in Fig.~\ref{fig:stcav}a–c and Fig.~\ref{fig:xtcav}a–c. 
For each configuration, between 100 and 250 single-shot measurements were recorded both early in the accelerator using the S-band TCAV and downstream of the undulators using the X-band TCAV.
The corresponding averaged current projections and representative spectrograms are shown for the upstream S-band TCAV (Fig.~\ref{fig:stcav}d–f) and downstream X-band TCAV measurements in Fig.~\ref{fig:xtcav}d–f.

Faint horizontal and vertical fringe patterns visible in several S-band TCAV images arise from partial overlap of the electron beam with a fixed calibration mark on the deflector screen for a subset of shots. 
Because these fringes are stationary in screen coordinates and not correlated with the beam’s energy–time phase space, they do not affect interpretation of the projected current profiles.
The features were retained in the raw data to avoid introducing bias through aggressive filtering or smoothing that could remove genuine beam structure.

The S-band TCAV phase-space maps (Fig.~\ref{fig:stcav}) already exhibit localized regions of intensity consistent with the temporal spacing of the UV subpulses.
However, at this low-energy injector stage the longitudinal phase space is dominated by strong correlated energy chirp arising from RF curvature and space-charge forces early in the accelerator. 
These dynamics rotate density modulations into energy space and produce substantial overlap of temporal slices in the spectrometer.
Combined with the limited energy resolution of the spectrometer, this prevents direct resolution of the intrinsic slice energy spread and obscures fine-scale temporal structure in the projected current profiles.
The apparent energy broadening in Fig.~\ref{fig:stcav} therefore reflects correlated phase-space evolution rather than true slice energy spread.

After acceleration through the linac sections and compression in BC1 and BC2, the laser-imprinted structure becomes clearly resolvable in the X-band TCAV (Fig.~\ref{fig:xtcav}).
While the imposed laser shaping establishes the initial temporal structure of the electron bunch, strong space-charge forces in the injector region lead to nonlinear evolution of the current profile, including phase-space rotation and redistribution of charge, such that the downstream profiles represent transformed rather than direct replicas of the drive laser. 
Longitudinal dispersion in the chicanes amplifies initial energy–time correlations, while acceleration to ultra-relativistic energies and collective effects--including space charge, RF curvature, coherent synchrotron radiation, and wakefields--further enhance energy modulation and phase-space separation. 
Although the X-band TCAV provides finer intrinsic temporal streaking resolution than the S-band system, the primary improvement in downstream visibility arises from this amplified energy–phase-space separation rather than temporal resolution alone.

Among the tested configurations, the deep three-hump modulation (Distribution 2; Fig.~\ref{fig:xtcav}b) produced the most distinct structure, showing three localized hot spots and corresponding peaks in the current projection (Fig.~\ref{fig:xtcav}e and~\ref{fig:xtcav}h). 
The deep two-hump (distribution 3; Fig.~\ref{fig:xtcav}c) case displayed two bright lobes, though a strong residual energy chirp rotated the phase space such that the bimodal structure merged in the current projection but is more visible in an energy spectrum projection (Fig.~\ref{fig:xtcav}f and~\ref{fig:xtcav}i). 
Similarly, the shallow three-hump profile (distribution 1; Fig.~\ref{fig:xtcav}a) exhibited weaker but discernible correlations between UV subpulses and downstream modulations on average (Fig.~\ref{fig:xtcav}d); however, on a shot-to-shot basis the full spectrograms indicate a strong triple hump structure (Fig.~\ref{fig:xtcav}g). 
While further tuning could have optimized visibility, beamtime constraints limited the extent of parameter scans.

Such projection-dependent visibility is consistent with phase-space rotation and compression dynamics at high peak current~\cite{Behrens2014, PhysRevLett.102.254801}.
While numerical de-chirping could further sharpen the projected profiles, only centroid alignment and normalization were applied here to preserve the raw phase-space statistics.
Collectively, these measurements confirm that programmable temporal shaping of the photoinjector laser imprints measurable structure on the electron bunch and that this structure is amplified and preserved through the full accelerator chain. The data processing steps are further discussed in Methods, Section~\ref{sec:methodsD}.

\begin{figure}[hpbt]
\centering
\includegraphics[width=\linewidth]{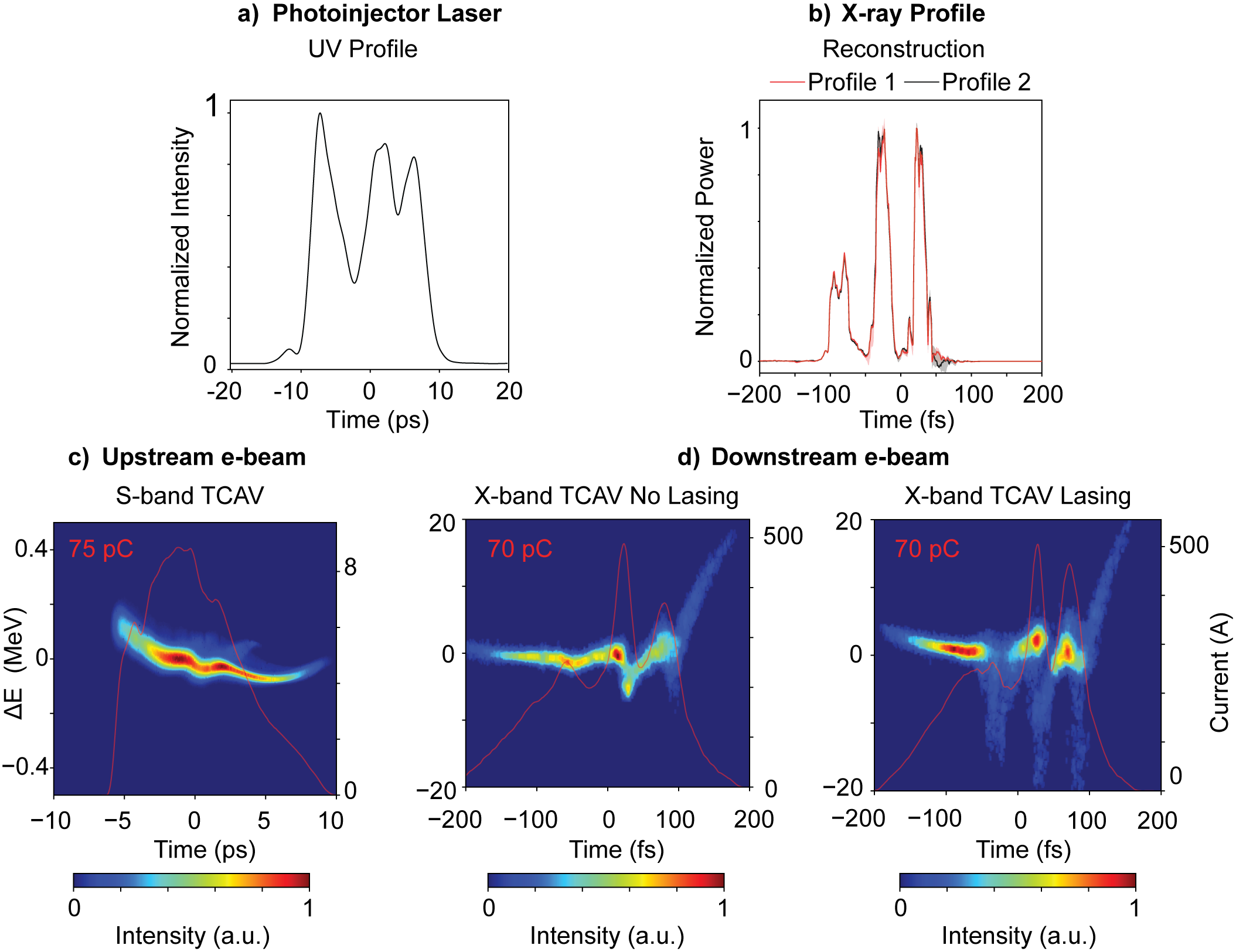}
\caption{
\textbf{UV to X-ray temporal structure.}
(a) Measured ultraviolet photoinjector laser temporal profile for the deep three-hump modulation (distribution 2). 
(b) Reconstructed X-ray temporal power profiles obtained using the X-band TCAV energy-loss method with two independently matched lasing-off reference shots (Profiles 1 and 2). Shaded regions indicate reconstruction variability obtained by shifting the lasing-off phase-space image by small integer time bins relative to the lasing-on shot. 
(c) Representative upstream S-band TCAV measurement of the electron bunch immediately after the photoinjector. 
(d) Downstream X-band TCAV phase-space measurements after the undulator for lasing-off and lasing-on conditions.
}
\label{fig:uv_to_xray}
\end{figure}

Having established that the imposed laser modulation survives acceleration and compression, we next examine how this structured electron beam influences the emitted X-ray pulse. 
To probe how these modulations influence lasing, we tracked the deep triple-hump (distribution 2) configuration through the post-undulator X-band TCAV with the FEL lasing alternately disabled and enabled under similar conditions (Fig.~\ref{fig:uv_to_xray}). 
With lasing off, three distinct hot spots appear in the longitudinal phase space after compression, corresponding to the UV subpulses.
When lasing is on, these features persist but exhibit depletion due to energy extraction and microbunching in the undulators. 
The slice-dependent energy loss forms valley-like depletion bands characteristic of active lasing.

These depletion regions were used to perform an approximate X-ray temporal reconstruction (Fig.~\ref{fig:uv_to_xray}b).
A representative lasing-on shot was paired with lasing-off reference shots selected from an ensemble of no-lasing measurements using a two-stage matching procedure (see Methods, Section~\ref{sec:methodsE}). 
Candidate references were first restricted to those whose bunch charge matched the lasing-on shot within $\pm$3 pC. 
The remaining candidates were then ranked using a weighted correlation metric comparing both the longitudinal current profile and the slice-dependent center-of-mass (COM) energy profile within the region of significant beam current.
The two highest-ranked references are shown as Profile 1 and Profile 2 in Fig.~\ref{fig:uv_to_xray}b.

Following standard X-band TCAV analysis procedures~\cite{Behrens2014, Lane2025XTCAVDoc}, matched lasing-on and lasing-off phase-space measurements were used to reconstruct the relative X-ray temporal power profile (see Methods, Section~\ref{sec:methodsE}). 
In this work we employ a variance-based reconstruction approach in which the slice energy spread $\sigma_E$ is extracted from the second moment of the measured phase-space distribution and compared between lasing-on and lasing-off conditions. 
The resulting difference in slice energy spread, combined with the lasing-on current profile, provides an estimate of the relative X-ray power versus time. 
This $\sigma_E$-based reconstruction is commonly used as an alternative to the center-of-mass energy-loss method and can provide improved stability for beams with strong current modulation.

The resulting reconstruction exhibits three principal peaks on the compressed femtosecond timescale, broadly consistent with the structured modulation imposed on the drive laser. 
% Because TCAV reconstructions depend on both the choice of lasing-off reference shot and the relative temporal alignment between lasing-on and lasing-off measurements, the detailed fine structure should be interpreted as approximate~\cite{Goetzke:25}. 
Because TCAV reconstructions depend on both the choice of lasing-off reference shot and the relative temporal alignment between lasing-on and lasing-off measurements, the detailed fine structure should be interpreted as approximate~\cite{Goetzke:25}. 
To assess this sensitivity, we evaluated the reconstruction while shifting the lasing-off reference image by small integer time bins around the optimal alignment. 
The shaded regions in Fig.~\ref{fig:uv_to_xray}b show the range of reconstructed profiles obtained by scanning the relative alignment by $\pm10$ temporal bins, reflecting the uncertainty associated with determining the precise temporal overlap between lasing-on and lasing-off phase-space measurements. 
This alignment scan provides a practical estimate of the reconstruction uncertainty arising from shot matching and temporal registration.
To assess this sensitivity, we evaluated the reconstruction while shifting the lasing-off reference image by small integer time bins around the optimal alignment. 
% The shaded regions in Fig.~\ref{fig:uv_to_xray}b show the range of reconstructed profiles obtained by scanning the relative alignment by $\pm10$ temporal bins, reflecting the uncertainty associated with determining the precise temporal overlap between lasing-on and lasing-off phase-space measurements. 
% While the precise amplitudes of individual peaks vary slightly within this range, the multi-peaked envelope persists across both reference shots and across nearby alignment shifts, indicating that the underlying temporal modulation is robust.

Importantly, the structured laser distributions used here produce a more complex reconstruction scenario than the single-pulse or weakly modulated cases typically analyzed with TCAV-based reconstruction methods. 
The shaped photoinjector drive laser generates multiple closely spaced current spikes, each capable of contributing to FEL energy extraction. 
In this regime the reconstructed X-ray profile reflects the combined influence of the modulated current distribution, the evolving energy chirp, and collective beam dynamics within the accelerator. 
While wakefields and shot-to-shot fluctuations introduce additional uncertainty in the precise relative amplitudes of individual peaks, the consistent correspondence between the spacing of the reconstructed features and the imposed UV modulation supports the conclusion that source-level laser shaping influences the temporal structure of the emitted X-ray pulse.

Comparing no-lasing and lasing conditions highlights this connection directly: with lasing active, the slice energy spread increases near the modulated current spikes, and the reconstructed X-ray profile exhibits a similar temporal structure. 
Although limited machine tuning constrained full optimization, similar structured phase-space signatures were observed across multiple tested drive-laser configurations, suggesting that the observed behavior is not unique to a single shot pairing.
Together, these observations indicate a pathway by which programmable photoinjector laser shaping can influence downstream electron phase space and, in turn, the temporal structure of the emitted X-ray pulse. 
This represents an initial experimental step toward laser-informed control of X-ray pulse shaping and marks the first demonstration of laser-programmed X-ray pulse shaping at the LCLS-II photoinjector, establishing a foundation for future adaptive control of free-electron-laser output.

\section{Conclusions}
We have demonstrated, for the first time, longitudinal programmable shaping of free-electron laser X-ray pulses via direct modulation of the ultraviolet photoinjector drive laser beyond the flattop profile. 
By integrating dispersion-controlled nonlinear synthesis with upstream programmable spectral shaping using a spatial light modulator, we establish a source-level control architecture that links laser waveform design to electron-beam longitudinal phase space and, ultimately, to the temporal structure of the emitted X-rays.

Through a series of controlled ultraviolet temporal profiles implemented at the LCLS-II photoinjector, we show that laser-imposed structure is preserved through acceleration and compression, amplified by longitudinal dispersion, and imprinted on the X-ray emission. 
Time-resolved measurements using transverse deflecting cavities before and after the undulators provide direct experimental evidence of this laser-to-electron-to-X-ray mapping, confirming that programmable photoinjector shaping constitutes a viable actuator for high-repetition-rate FEL control.

These results establish a foundation for adaptive, source-level programmability in next-generation FEL facilities. 
In the MHz regime, where data volumes preclude exclusive reliance on offline analysis, fast and flexible control knobs at the electron source become essential. 
Programmable photoinjector laser shaping provides such a knob, enabling deterministic modification of the electron beam’s initial longitudinal phase space on experimentally relevant timescales.

When combined with start-to-end modeling, high-throughput diagnostics, and emerging real-time optimization frameworks, this source-level programmability enables a clear pathway toward adaptive and ultimately autonomous FEL operation. 
More broadly, the demonstrated laser-to-electron-to-X-ray mapping opens new opportunities for real-time optimization, multiplexed operation, and dynamic tailoring of X-ray pulse structures, advancing precision control of ultrafast X-ray sources.

\section{Methods}
\label{sec:methods}
\subsection{Spatial Light Modulator–Based Spectral Shaping}
\label{sec:methodsA}
Programmable control of the photoinjector laser waveform was implemented using a liquid-crystal spatial light modulator (LC-SLM) configured in a folded 4f spectral shaping geometry. 
The choice of an LC-SLM was guided by several practical constraints specific to high-repetition-rate photoinjector laser operation. 
While acousto-optic programmable dispersive filters (AOPDFs) such as Dazzler systems offer high-speed spectral phase and amplitude modulation, current commercially available devices are not compatible with continuous operation at the full repetition rate of the photoinjector laser system and impose additional limitations on optical efficiency and pulse energy. 
In contrast, an LC-SLM provides static or quasi-static programmability with high spectral resolution, high damage threshold, and near-unity duty cycle, making it well suited for proof-of-concept operation within the available laser energy and efficiency margins.

The spectral shaper was built around a Santec SLM-210 reflective liquid-crystal modulator (1920 $\times$ 1200 pixels, 15-mm active length), operated in a folded 4f geometry using an Ibsen PING-1379-422 transmission grating (1379 grooves/mm) and a concave spherical mirror with a focal length of 100 mm. 
The input infrared pulse was centered at 1035 nm with a total spectral bandwidth of approximately 70 nm (16 nm FWHM).
The grating dispersion and focal length were chosen such that approximately 76 nm of optical bandwidth was mapped across the full horizontal extent of the SLM, providing near-uniform spectral coverage while slightly underfilling the active area. 
In this configuration, individual wavelength components were spatially dispersed and mapped to distinct columns of SLM pixels, enabling deterministic spectral phase control through column-wise modulation.

The LC-SLM operates by electrically tuning the birefringence of each pixel, thereby imposing a programmable phase delay on the reflected optical field. 
Phase masks were written as two-dimensional grayscale images, with the horizontal pixel coordinate corresponding to optical frequency and the vertical coordinate held uniform. 
For the experiments reported here, spectral phase profiles were parameterized using low-order polynomial functions mapped directly onto the SLM columns, enabling smooth and reproducible phase shaping while minimizing high-spatial-frequency phase discontinuities that could introduce unwanted spatiotemporal coupling.

The system was initially designed to support independent spectral phase and amplitude control using a polarization-based architecture in which two halves of the SLM would modulate orthogonal polarization components separated and recombined using a Wollaston prism. 
However, to reduce complexity and maximize operational robustness during the limited beamtime, the implementation was simplified to operate primarily in a phase-only shaping mode. 
Limited amplitude shaping was nonetheless achieved by selectively blanking contiguous spectral regions on the SLM--setting the corresponding pixel columns to zero phase retardance--thereby suppressing the contribution of those frequency components in the recombined pulse.
While this approach does not provide continuous amplitude control, it enables coarse spectral filtering sufficient for proof-of-concept waveform tailoring.

Initial calibration and baseline operation were established by compensating dispersion introduced by insertion of the SLM-based shaping optics.
First, the native oscillator pulse was characterized directly using a home-built second-harmonic-generation frequency-resolved optical gating (SHG-FROG) diagnostic placed upstream of the shaping system, providing a reference spectrogram. 
The SLM was then inserted with a flat phase mask (zero applied phase), and the transmitted pulse was measured again with the same SHG-FROG system, revealing residual dispersion introduced by the 4f shaper optics. 
Initial estimates of second- and third-order dispersion were applied to the SLM phase mask based on visual comparison of the spectrograms, followed by an iterative error-minimization routine that refined the spectral phase coefficients to minimize the difference between the reference and post-SLM FROG measurements. 
This optimized phase mask defined the baseline operating point of the shaping system, ensuring faithful reproduction of the unshaped oscillator pulse prior to applying structured waveform modulation.

\subsection{UV Shaping \& Measurements with XCorr}
\label{sec:methodsB}
Ultraviolet pulse shaping was implemented using the DCNS technique, which enables programmable temporal waveform generation through nonlinear frequency conversion under controlled dispersion.
In DCNS, the infrared pulse is split into two branches--a dispersive stretcher and a dispersive compressor--implemented using off-Littrow grating geometries that impart opposite-sign second- and third-order dispersion (SOD and TOD). 
A chirped volume Bragg grating (CVBG) is used in combination with these grating-based lines to adjust the absolute magnitude of the dispersion, while preserving the intrinsic sign inversion between the two branches.
The combined action of the stretcher, compressor, and CVBG defines the net SOD and TOD of the synthesized UV pulse. 
A complete theoretical and experimental description of the DCNS framework is provided in Ref. Lemons, et al~\cite{lemons2025nonlinear}.

In our implementation, both the stretcher and compressor were static optical lines, providing a stable and reproducible dispersion environment during beamtime. 
A weak spectral amplitude mask was introduced in the stretcher branch, where the spectrum is spatially dispersed. 
This mask was used primarily to suppress poorly converted frequency components near the edges of the available bandwidth. 
The CVBG employed in the system was not perfectly centered on the laser spectrum, resulting in partial clipping of the spectral wings; selectively removing these components improved the smoothness and reproducibility of the resulting UV temporal profile. 
While the amplitude mask also enabled limited additional shaping flexibility, its dominant role in this work was spectral cleanup rather than precise amplitude modulation.

Removal of these edge frequencies was observed to produce smoother UV temporal envelopes. 
This behavior is consistent with the nonlinear nature of DCNS, in which spectral components that are strongly attenuated in one branch do not contribute efficiently to the nonlinear synthesis process and can otherwise introduce temporal artifacts in the synthesized pulse. 
For the representative UV pulse shapes presented here, the required dispersion values were on the order of $\pm$2.56 ps$^2$ (SOD) and $\pm$0.28 ps$^3$ (TOD).
These dispersion values were realized entirely through static optical elements, ensuring stability over the duration of individual measurements.

Direct temporal characterization of the long-duration UV pulses generated by DCNS remains experimentally challenging. 
Although recent advances in XFROG-based techniques have extended access to this regime, for the present work, we employed a cross-correlation (XCorr) measurement using an APE pulseCheck system configured in difference-frequency generation (DFG) mode.
In this configuration, the shaped UV pulse was mixed with the unshaped infrared oscillator pulse in a collinear geometry, providing a direct measurement of the UV temporal envelope. 
Because the photocathode response depends primarily on the temporal profile of the drive laser, this diagnostic was sufficient for evaluating longitudinal shaping of the electron bunch.

The XCorr was operated with a photomultiplier tube (PMT) detector to accommodate the low UV pulse energies available at this stage of the system. 
According to the manufacturer's specifications, the PMT detector requires operation at repetition rates in the tens-of-kilohertz range to achieve adequate signal-to-noise performance, with external triggering supported from 10 Hz up to 50 kHz. 
This requirement introduced a practical mismatch during beamtime, as the accelerator experiment itself was conducted at approximately 120 Hz.

To address this constraint, UV pulse-shape measurements were performed at a repetition rate of approximately 25 kHz. 
At this rate, the DCNS process in the BBO crystal remained sufficiently stable, and no significant thermal distortion of the UV temporal profile was observed. 
At higher repetition rates, thermal effects in the nonlinear crystal led to noticeable changes in pulse shape; under routine operation, this would be mitigated by establishing a stable baseline and shaping relative to that operating point. 
For the present proof-of-principle experiment, UV measurements at 25 kHz provided a practical compromise that enabled reliable diagnostics without introducing substantial thermal artifacts. 
Averaging was employed to suppress aliasing effects.

Over longer timescales, slow beam-pointing drift was observed, attributable primarily to the mechanical stability of the grating mounts in the prototype DCNS line rather than the intrinsic instability of the shaping process. 
This drift occurred on timescales well separated from individual measurements and did not limit the proof-of-principle results reported here.
Subsequent iterations of the system will incorporate improved optomechanical supports and active stabilization to enable robust operation over extended experimental runs.

Confidence in the inferred UV temporal profiles was further supported by consistency between the cross-correlation measurements and independent diagnostics of the resulting electron-bunch longitudinal profiles downstream of the photoinjector. 
While future experiments will incorporate more advanced UV pulse characterization techniques, the combination of DCNS-based shaping, cross-correlation measurements, and electron-beam verification was sufficient to demonstrate programmable control of the photoinjector drive laser.

\subsection{Beamline Setup \& Diagnostics}
\label{sec:methodsC}
\subsubsection{Photoinjector Laser Delivery and Alignment}
The UV drive laser is generated from a Yb-based chirped-pulse-amplification (CPA) system consisting of a Light Conversion Flint oscillator seeding a Carbide regenerative amplifier. 
The laser system is phase-locked to the accelerator radio-frequency reference, providing sub–100-fs timing stability between the laser and the electron beam. 
After programmable temporal shaping and nonlinear upconversion to the ultraviolet (described in the preceding subsection), the UV beam is transported from the laser room to the injector tunnel through evacuated beamlines to preserve pulse fidelity and pointing stability.

At the photocathode, a spatially approximate flattop transverse profile was produced by relay-imaging an iris aperture, which clipped the Gaussian wings of the beam and transmitted only the central plateau.
This ensured a well-defined transverse distribution while allowing the longitudinal electron-bunch profile to be controlled exclusively through temporal shaping of the UV pulse. 
Laser alignment and spot-size verification were performed using standard upstream laser diagnostics and confirmed in situ using downstream electron-beam imaging.
% The measured UV laser transverse profile corresponding to this configuration is shown in Supplementary Fig.~S1.

\subsubsection{Photoinjector Beamline Configuration and Diagnostics}
The experiments were performed at the LCLS-II soft X-ray (SXR) beam line, which is based on a superconducting RF (SRF) linac capable of continuous-wave (CW) operation at MHz repetition rates. 
Electron bunches are generated in a normal-conducting very-high-frequency (VHF) photoinjector gun, where the shaped UV laser pulse illuminates the photocathode to initiate electron emission. 
The gun is followed by a buncher cavity and a set of focusing solenoids that provide initial longitudinal compression and transverse matching of the beam. The LCLS-II injector design and operating principles are described in detail in Ref.~\cite{NEVEU2025170065}.

Downstream of the injector, a series of superconducting cryomodules accelerate the beam to energies of up to approximately 4 GeV.
Two magnetic bunch compressors (BC1 and BC2), located after the L1 and L2 linac sections, are used to tailor the longitudinal phase space and achieve kiloampere-class peak currents when required. 
The beam is subsequently transported to the SXR undulator line, which produces femtosecond X-ray pulses spanning photon energies from approximately 0.25 to 5 keV.

\subsubsection{Electron Beam Diagnostics}
Transverse beam properties were monitored using scintillator and optical transition radiation (OTR) screens located along the injector and early linac sections. 
A downstream Ce:YAG screen imaged with a gated camera was used to record the transverse electron-beam profile. 
Transverse emittance measurements were obtained using the standard quadrupole-scan technique, in which the strength of an upstream quadrupole magnet is varied and the resulting beam size is measured at a downstream screen. 

Longitudinal phase-space diagnostics were provided by radio-frequency transverse deflecting cavities (TCAVs) positioned at multiple locations along the beamline.
An S-band TCAV located after accelerator section L0 enables measurements of the early longitudinal phase space and injector-level bunch structure, while an X-band TCAV installed downstream of the undulator lines provides higher-resolution diagnostics of the compressed beam. 
These systems map the temporal coordinate of the bunch onto a transverse screen dimension, allowing single-shot measurements of the longitudinal current profile and energy–time correlations. 
However, since the TCAV diagnostic is destructive of the electron beam, simultaneous single-shot S-band and X-band measurements are not possible. 
A detailed description of TCAV calibration procedures, temporal resolution, and data processing is provided in Methods, Section~\ref{sec:methodsD}.

Shot-by-shot X-ray pulse energy was monitored using gas-monitor detectors (GMDs), which measure photoionization signals from rare gases at low pressure.
All diagnostic data were acquired using the facility data-acquisition system and synchronized to the accelerator timing and laser settings.

\subsection{TCAV Data Processing}
\label{sec:methodsD}
Longitudinal phase-space measurements of the electron beam were performed using radio-frequency TCAVs, which map the temporal structure of the bunch onto a transverse coordinate at a downstream screen. 
In this work, both S-band and X-band TCAV systems available at LCLS-II were used, following standard operating and calibration procedures established for these diagnostics~\cite{dolgashev24lcls, dolgashev2022new}.
Here the measured energy axis reflects the full correlated energy–time phase space of the bunch; intrinsic slice energy spread would require dechirping and localized projections, which were not applied in this work.

The time calibration begins by converting RF phase to time using the streaking cavity wavelength $\lambda_s$ and the speed of light $c$. 
The temporal increment per degree of RF phase is given by
\begin{equation}
t_{\deg} = \frac{\lambda_s}{360\,c} \quad [\mathrm{s/deg}],
\end{equation}
where $\lambda_s = c / f_s$ and $f_s$ is the cavity frequency. 
To convert screen pixel units into femtoseconds, this factor is combined with the measured screen resolution $\mathrm{res}$ (in $\mu$m/pixel) and the streaking strength $\mathrm{streak}$ (in $\mu$m/deg), yielding
\begin{equation}
\mathrm{px\;to\;fs} = \frac{t_{\deg}\ \times \mathrm{res}}{\mathrm{streak}} \times 10^{15}.
\end{equation}

The streaking strength was obtained from phase scans of the TCAV, where the beam centroid displacement on the screen was measured as a function of RF phase. 
Screen resolution and pixel calibration were extracted from profile-monitor metadata files.

The energy axis is determined independently using the known vertical dispersion at the screen downstream of a dipole magnet. 
Pixel units are converted to energy via
\begin{equation}
\mathrm{px\;to\;MeV} = \frac{\mathrm{res}}{\mathrm{dispersion}} \times E_0 ,
\end{equation}
where $\mathrm{dispersion}$ is the vertical dispersion in $\mu$m/MeV and $E_0$ is the beam energy. 
This calibration accounts for the finite unstreaked beam size, which ultimately limits the resolvable energy separation.

Raw TCAV images were further processed using a custom analysis pipeline designed to robustly extract longitudinal current profiles while rejecting frames affected by noise, background fluctuations, or partial beam clipping. 
Prior to analysis, images were optionally background-subtracted using averaged dark frames acquired under identical camera and beamline conditions, converted to floating-point representation, and lightly filtered to suppress isolated pixel noise while preserving the beam envelope.

Morphological filtering was then performed using standard grayscale operations implemented with OpenCV.
A light erosion step was applied using a 3$\times$3 square structuring element and a single iteration, in which each output pixel was set to the minimum intensity within its local neighborhood while maintaining the original image dimensions. 
This operation locally contracts bright regions by approximately one pixel, effectively removing isolated hot pixels and thin background artifacts without distorting the multi-pixel-wide beam distribution.

An implicit segmentation mask was then applied via a threshold-to-zero operation, where pixels with intensities below the adaptive threshold were set to zero while pixels above threshold retained their original values. 
Finally, a grayscale dilation step using the same 3$\times$3 structuring element and one iteration was applied to the thresholded image, replacing each pixel with the local maximum intensity to re-expand the surviving beam region, fill small gaps, and smooth the beam boundary.

The threshold value was determined dynamically for each dataset using robust background statistics computed from the image ensemble. 
Specifically, the median intensity and median absolute deviation (MAD) were used to estimate the background level, with the threshold defined as a fixed multiple of the MAD above the median.
This approach enabled stable rejection of background noise while remaining insensitive to outliers and shot-to-shot fluctuations. 
In cases where the MAD was ill-defined, a conservative fallback threshold based on the 95th percentile of the background intensity distribution was employed.

Following filtering, each image was evaluated for physical validity prior to inclusion in further analysis.
A projection along the temporal axis was computed from the uncentered image, and edge regions were examined to detect truncated or clipped beams. 
Frames exhibiting anomalously large signals near the image boundaries--indicative of incomplete capture of the longitudinal phase space--were automatically excluded. 
This procedure ensured that only fully contained bunch profiles contributed to the final dataset.

For accepted shots, the filtered image was normalized by its total intensity and scaled by the measured bunch charge.
The beam centroid was then determined using an intensity-weighted average, and the image was recentered accordingly. 
No additional peak-alignment procedures were applied; all profiles were analyzed in their native temporal reference frame to preserve physically meaningful timing information.

Longitudinal current profiles were obtained by projecting the centered images onto the time axis.
To suppress high-frequency noise while preserving femtosecond-scale structure, the projections were smoothed using a moving-average window matched to the experimental temporal resolution.
Energy projections were computed analogously and used as a diagnostic for correlated energy spread and beam quality.

For quantitative analysis, only images passing all quality and clipping criteria were retained.
Shot-to-shot fluctuations were reduced by averaging over multiple valid acquisitions.
The full processing chain--background subtraction, thresholding, clipping rejection, normalization, and projection--was applied identically to all datasets used in this study.

All TCAV measurements were performed near the RF zero-crossing, ensuring a linear mapping between longitudinal time and transverse momentum. 
Calibration parameters, including streaking strength, dispersion, and pixel resolution, were obtained from dedicated calibration scans and remained stable over the duration of each experimental shift. 
% A complete list of processing parameters, thresholds, and implementation details required for reproducibility is provided in the Supplementary Material.

\subsection{XTCAV Shot Matching and X-ray Temporal Reconstruction}
\label{sec:methodsE}

An approximate single-shot X-ray temporal profile was reconstructed for the Distribution 2 configuration using post-undulator X-band TCAV measurements following established FEL analysis techniques~\cite{Behrens2014, Lane2025XTCAVDoc}. 
A representative lasing-on phase-space image was selected and paired with a lasing-off reference from an ensemble of baseline measurements acquired with FEL interaction suppressed via orbit perturbation. 
A two-stage matching procedure was applied to identify a suitable reference shot.

First, lasing-off candidates were filtered to match the bunch charge of the selected lasing-on shot within $\pm$3 pC. 
Among these charge-matched shots, both the longitudinal current profile and the slice-dependent center-of-mass (COM) energy profile were extracted from the processed TCAV images and compared to those of the lasing-on shot. 
Similarity was quantified using a normalized correlation metric evaluated within the temporal region of significant beam current. 
The final matching score combined correlations of the COM-energy and current profiles with weights of 0.65 and 0.35, respectively. 
The lasing-off shot with the highest similarity score was selected as the reference for reconstruction, while the next-best match was retained as a secondary comparison.

Both lasing-on and lasing-off images were processed using identical analysis steps, including background subtraction, threshold filtering, and conversion from pixel coordinates to physical time and energy axes using measured TCAV calibration parameters. 
The longitudinal current profile $I(t)$ was obtained from the horizontal projection of the normalized phase-space image and scaled using the measured bunch charge.

In addition to the slice energy centroid, the slice energy spread was extracted from the second moment of the phase-space distribution,
\begin{equation}
\sigma_E^2(t) =
\frac{\int (E-\langle E(t)\rangle)^2\,\rho(E,t)\,dE}
{\int \rho(E,t)\,dE},
\end{equation}
where $\rho(E,t)$ denotes the processed TCAV intensity distribution.

Two common approaches are used to estimate FEL power from TCAV measurements: a center-of-mass energy-loss method based on slice energy depletion and a variance-based method based on changes in slice energy spread. 
In this work we employ the variance-based reconstruction, which can provide improved stability for beams with strong current modulation. 
The FEL-induced increase in slice energy spread between lasing-on and lasing-off conditions was therefore used to estimate the relative X-ray power profile,
\begin{equation}
P(t) \propto
\left[
\sigma_{\mathrm{on}}^{2}(t) - \sigma_{\mathrm{off}}^{2}(t)
\right]
\, I_{\mathrm{on}}^{2/3}(t),
\end{equation}
where $I_{\mathrm{on}}(t)$ denotes the lasing-on current profile.

To suppress noise-dominated regions, the reconstruction was evaluated only within the temporal window defined by significant beam current. 
The resulting power profile was normalized so that its temporal integral matched the independently measured X-ray pulse energy obtained from shot-resolved gas detector measurements,
\begin{equation}
\int P(t)\,dt = E_{\mathrm{X\text{-}ray}} .
\end{equation}

To assess sensitivity to residual temporal alignment errors, the selected lasing-off reference was additionally shifted by small integer time bins relative to the lasing-on shot and the reconstruction repeated. 
Scanning the alignment by $\pm10$ temporal bins around the optimal overlap produced a family of reconstructed profiles that was used to estimate the uncertainty associated with temporal registration of the TCAV measurements. 
This procedure yields an approximate temporal envelope of the emitted X-ray pulse derived from FEL-induced changes in the electron-beam phase-space distribution.

%%%%%%%%%%%%%%%% REFERENCES %%%%%%%%%%%%%%%

\clearpage % Clear all remaining figures and tables then start a new page

% The list of references goes after the main text and before the acknowledgements
% When preparing an initial submission, we recommend you use BibTeX, like this:
%
\bibliography{science_template} % for a file named science_template.bib
\bibliographystyle{sciencemag}

%%%%%%%%%%%%%%%% ACKNOWLEDGEMENTS %%%%%%%%%%%%%%%

\section*{Acknowledgments}
The authors would like to acknowledge Mike Dunne for his support in enabling the beamtime during which these experiments were conducted. His advocacy was instrumental in securing the necessary equipment and ensuring sufficient facility access to carry out the studies presented here.
The authors would also like to acknowledge Gregory Stewart for his help with graphic editing.
\paragraph*{Funding:}
The authors acknowledge the support from the SLAC National Accelerator Laboratory, the U.S. Department of Energy (DOE), the Office of Science, Office of Basic Energy Sciences under Contract No. DE-AC02-76SF00515, No. DE-SC0022559, No. DE-FOA-0002859, the National Science Foundation under Contract No. 2231334 and 2436343, and the U.S. Department of Defense via AFOSR Contract No. FA9550-23-1-0409 and the National Defense Science and Engineering Graduate Fellowship. 
Additionally, Kurtis Borne was supported by the Chemical Sciences, Geosciences and Biosciences Division, Office of Basic Energy Sciences, Office of Science, US Department of Energy, grant No. DE-FG02-86ER13491.

% \paragraph*{Author contributions:}
% List each author’s contributions to the paper.
% Use initials to abbreviate author names.
% \paragraph*{Competing interests:}
% There are no competing interests to declare.
% \paragraph*{Data and materials availability:}

\end{document}